\documentclass[10pt, conference, compsocconf]{IEEEtran}

\usepackage[cmex10]{amsmath}
\usepackage[dvips]{graphicx}
\usepackage{fancyvrb}
\usepackage{cite}
\usepackage{amsmath,zed-csp,alltt}
\usepackage{fancyvrb,booktabs,paralist}

\usepackage{makeidx,graphics}
\usepackage{graphicx,fancyhdr}

\renewcommand{\seq}{\mbox{~\large{;}}~}
\newcommand{\sq}{\mbox{~;~}}

\newcommand{\cpair}{\div}

\newcommand{\close}[1]{[~#1~]}

\begin{document}

\title{Web Service Composition - BPEL vs cCSP Process Algebra}

\author{\IEEEauthorblockN{Shamim Ripon, Mohammad Salah Uddin and Aoyan Barua}
\IEEEauthorblockA{Department of Computer Science and Engineering\\
East West University,
Dhaka, Bangladesh\\
Email: dshr@ewubd.edu}}

\maketitle

\begin{abstract}
Web services technology provides a platform on which we can develop distributed services. The interoperability among these services is achieved by various standard protocols. In recent years, several researches suggested that process algebras provide a satisfactory assistance to the whole process of web services development. Business transactions, on the other hand, involve the coordination and interaction between multiple partners. With the emergence of web services, business transactions are conducted using these services. The coordination among the business processes is crucial, so is the handling of faults that can arise at any stage of a transaction. BPEL models the behavior of business process interaction by providing a XML based grammar to describe the control logic required to coordinate the web services participating in a process flow. However BPEL lacks a proper formal description where the composition of business processes cannot be formally verified. Process algebra, on the other hand, facilitates a formal foundation for rigorous verification of the composition. This paper presents a comparison of web service composition between BPEL and process algebra, cCSP.
\end{abstract}


\section{Introduction}

Web services technology provides a platform on which we can develop distributed services. The interoperability among these services is achieved by the standard protocols (WSDL~\cite{wsdl2001}, UDDI~\cite{uddi}, SOAP~\cite{Curbera02}) that provide the ways to describe services, to look for particular services and to access services. With the emergence of web services, business transactions are conducted using these services~\cite{cacm03}. Web services provided by various organizations can be inter-connected to implement business collaborations, leading to composite web services.
Business collaborations require interactions driven by explicit process models. Web services are distributed, independent processes which communicate with each other through the exchange of messages. The coordination between business processes is particularly crucial as it includes the logic that makes a set of different software components become a whole system. Hence it is not surprising that these coordination models and languages have been the subject of thorough formal study, with the goal of precisely describing their semantics, proving their properties and deriving the development of correct and effective implementations.

Process calculi are models or languages for concurrent and distributed interactive systems. It has been advocated in~\cite{contracts, gwen:reasonWS} that process algebras provide a complete and satisfactory assistance to the whole process of web services development. Being simple, abstract, and formally defined, process algebras make it easier to formally specify the message exchange between web services and to reason about the specified systems. Transactions and calculi have met in recent years both for formalizing protocols as well as adding transaction features to process calculi \cite{express00, fsttcs03, bocchi03, join}.

Several research issues, both theoretical and practical, are raised by web services. Some of the issues are to specify web services by a formally defined expressive language, to compose them, and to ensure their correctness; formal methods provide an adequate support to address these issues~\cite{andrea:WS}. Recently, many XML-based process modeling languages (also known as choreography and orchestration~\cite{pletz:ieeews} languages) such as WSCI~\cite{wsci}, BPML~\cite{bpml}, BPEL4WS~\cite{BPEL}, WSFL~\cite{WSFL}, XLANG~\cite{XLANG} have emerged that capture the logic of composite web services. These languages also provide primitives for the definition of business transactions.

Several proposals have been made in recent years to give a formal definition to compensable processes by using process calculi. These proposals can be roughly divided into two categories. In one category, suitable process algebras are designed from scratch in the spirit of orchestration languages, e.g., BPEL4WS (Business Process Execution Language for Web Services, BPEL in short). Some of them can be found in~\cite{bruni05, StAC00, csp25}. In another category, process calculi like the $\pi$-calculus~\cite{milner:pi, parrow:pi} and the join-calculus ~\cite{cham-Join} are extended to describe the interaction patterns of the services where, each service declares the ways to be engaged in a larger process. Some of them are available in~\cite{bocchi03, join, webpi, Mazzara:Framework}.

Inspired by the growing interest in transaction processing using web services, this paper presents our on-going experiment of comparing the composition of web services by BPEL and process algebra cCSP~\cite{csp25}. cCSP is an extension of CSP, especially defined to model business transactions. The formal semantics of the algebra has already been defined~\cite{csp25, cCSP05}.We model the transaction of a Car Broker web service using both BPEL and cCSP examining the expressiveness of both languages.

In the remainder of the paper, Section~\ref{sec:cCSP}  gives an overview of the cCSP process algebra and its constructs. Section~\ref{sec:carbroker} first briefly describes our case study web service and then model the web service in both BPEL and cCSP. For brevity only an abstract version of the Car Broker web service is modeled in this paper. Finally, we conclude our paper and outline our future plans.

\section{Compensating CSP (cCSP)}\label{sec:cCSP}

\begin{figure}[!htb]
\centering
\includegraphics[width=.5\textwidth]{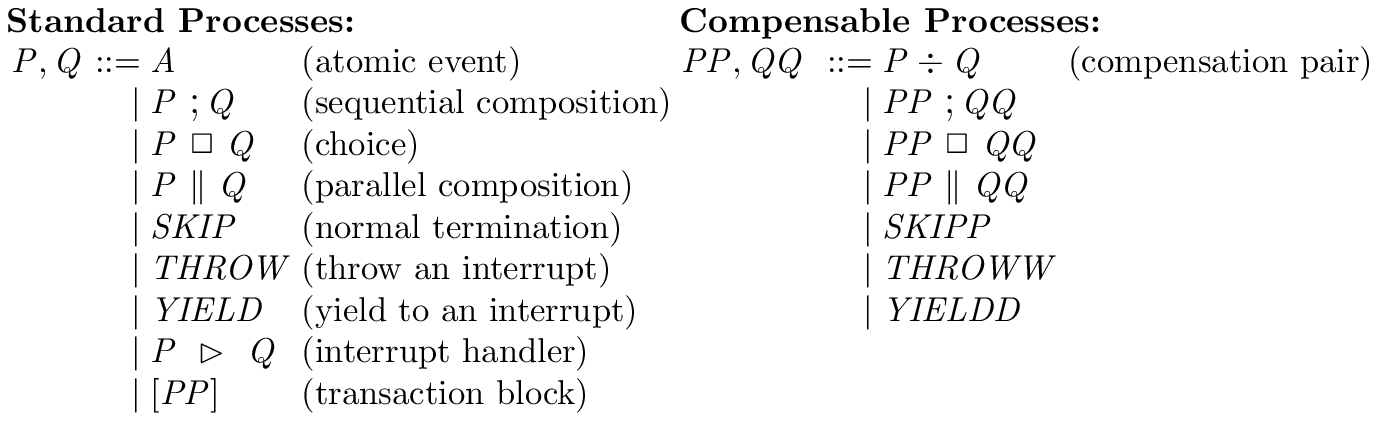}
\caption{cCSP syntax}
\label{fig:syntax}
\end{figure}

The introduction of the cCSP language was inspired by the combination of two ideas: transaction processing features, and process algebra. Like standard CSP, processes in cCSP are modeled in terms of the atomic events they can engage in. The language provides operators that support sequencing, choice, parallel composition of processes. In order to support failed transaction, compensation operators are introduced. The processes are categorized into standard, and compensable processes. A standard process does not have any compensation, but compensation is part of a compensable process that is used to compensate a failed transaction. We use notations, such as, $P,Q,..$ to identify standard processes, and $PP,QQ,..$ to identify compensable processes. A subset of the original cCSP is considered in this paper, which includes most of the operators. The cCSP syntax, considered in this paper, is summarized in  Fig.~\ref{fig:syntax}.

The basic unit of the standard processes is an atomic event ($A$). The other operators are the sequential~($P\sq Q$), and the parallel composition ($P\parallel Q$), the choice operator ($P\extchoice Q$), the interrupt handler ($P\rhd Q$), the empty process $SKIP$, raising an interrupt $THROW$, and yielding an interrupt $YIELD$. A process that is ready to terminate is also willing to yield to an interrupt. In a parallel composition, throwing an interrupt by one process synchronizes with yielding in another process. Yield points are inserted in a process through $YIELD$. For example, ($P\sq YIELD\sq Q$), is willing to yield to an interrupt in between the execution of $P$, and $Q$. The basic way of
constructing a compensable process is through a compensation pair ($P\cpair Q$), which is constructed from two standard processes, where $P$ is called the forward behaviour that executes during normal execution, and $Q$ is called the associated compensation that is designed to compensate the effect of $P$ when needed. The sequential composition of compensable processes is defined in such a way that the compensations of the completed tasks will be accumulated in reverse to the order
of their original composition, whereas compensations from the compensable parallel processes will be placed in parallel. In this paper, we define only the asynchronous composition of processes, where processes interleave with each other during normal execution, and synchronize during termination. By enclosing a compensable process $PP$ inside a transaction block $\close{PP}$, we get a complete transaction and the transaction block itself is a standard process. Successful completion of $PP$ represents successful completion of the block. But, when the forward behaviour of $PP$ throws an interrupt, the compensations are executed inside the block, and the interrupt is not observable from outside of
the block. $SKIPP, THROWW$, and $YIELDD$ are the compensable counterpart of the corresponding standard processes and they are defined as follows:
\begin{eqnarray*}
&& SKIPP = SKIP\cpair SKIP,\qquad YIELDD = YIELD\cpair SKIP\\
&& THROWW = THROW\cpair SKIP
\end{eqnarray*}

A cCSP process is described in terms of its interactions
with its environment or other processes. The interactions are
described by using atomic actions via channels as in standard
CSP. We add some constructs to the language as syntactic
sugar. A communication is an event described by a pair $c.v$
where $c$ is the name of the channel on which communication
takes place and v is the value of the message. A construct is
defined to allow an input of an item $x$ from a set $M$ to a channel
in and the value $x$ determines the subsequent behaviour. Output
is the complement of the input.

\begin{eqnarray*}
    in?x:M \seq Q(x) &~\defs~& \Extchoice_{x\in M}in.x\seq Q(x)\\
    out!x &~=~& out.x
\end{eqnarray*}

When drawing diagrams of processes, the channels are drawn
using arrows in the appropriate direction to define them as
input or output and labeled with channel name. Let $P$ and
$Q$ are two processes and $c$ an output channel of $P$ and an input channel of $Q$. When $P$ and $Q$ composed in parallel ($P\parallel
Q$), a communication $c.v$ can occur only when both processes
engage simultaneously, i.e., when $P$ outputs $v$ on the channel,
simultaneously $Q$ receives the value. The choice is a binary
operator. While modeling a transaction we use the indexed
version of the operator, which is defined as: $\Extchoice_{x~\in~S}P_x$, e.g.,
\begin{eqnarray*}
    \Extchoice_{x~\in~\set{S_1,S_2\ldots S_n}}P_x &~=~& P_{S_1}\extchoice P_{S_2}
    \ldots\extchoice P_{S_n}
\end{eqnarray*}

The parallel operator is associative. For processes $P$, $Q$ and $R$
\begin{eqnarray*}
P\parallel (Q \parallel R) &~~=~~& (P\parallel Q)\parallel R
\end{eqnarray*}

In the composition $P\parallel_X Q$, the processes synchronize over
events of the set $X$. We also use I/O parameters for compensation
pair:
\begin{eqnarray*}
	(A?x\cpair B.x) \seq P(x) &~=~&
	\Extchoice_{x\in S}~(A.x\cpair B.x)\seq P(x)
\end{eqnarray*}

\section{Car Broker Web Services}\label{sec:carbroker}

A car broker web service negotiates car purchases for its
buyers and arranges loans for these. The car broker uses two
separate web services: a Supplier to find a suitable quote
for the requested car model and a Lender to arrange loans.
Each web service can operate separately and can be used
in other web services. In this case study, our focus is on
how the processes communicate with each other and how the
compensations are handled when there is an interrupt. For
brevity, several details are abstracted from the description. The
original car broker example can be found in~\cite{turner05, turner07}.

\subsection{Broker Web Service}\label{sec:broker}

\begin{figure}[!htb]
\centering
\includegraphics[scale=.42]{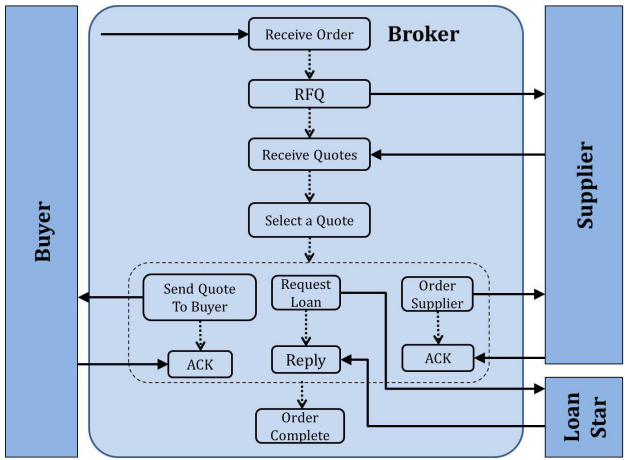}
\caption{Architectural view of Car Broker web Service}
\label{fig:broker}
\end{figure}

We model a car broker web service \textbf{Broker}. It provides
online support to customers to negotiate car purchases and
arranges loans for these. A buyer provides a need for a car
model. The broker first uses its business partner \textbf{Supplier} to
find the best possible quote for the requested model and then
uses another business partner \textbf{LoanStar} to arrange a loan for
the buyer for the selected quote. The buyer is also notified
about the quote and the necessary arrangements for the loan.
Both \textbf{LoanStar} and the \textbf{Buyer} can cause an interrupt to be invoked. A loan can be refused due to a failure in the loan
assessment and a customer can reject the loan and quoted
offer. In both cases, there is a need to run the compensation,
where the car might have already been ordered, or the loan
has already been offered. We first model this web service using BPEL and then in cCSP. The behaviour of the \textbf{Broker} we service in relation to BPEL modeling is illustrated in Fig~\ref{fig:broker}.

\begin{footnotesize}
\begin{verbatim}
<process name="CarBroker".../>
 <scope>
 <compensationHandler>
  <sequence>
  <invoke partnerLink="BrokerPL"
      operation ="cencelOrder"....../>
  </sequence>
  </compensationHandler>
 <sequence>
 <receive partnerLink="order_Broker",
    Variable="orderReq"...../>
  <scope>
  <compensationHandler>
	......
  </compensationHandler>
 <sequence>
 <invoke partnerLink="RFQ_Suppiler",
   outputVariable="SuppilerQuote",
   inputVariable="orderReq".../>
 <reply partnerLink="Quote_Broker",
  variable="SuppilerQuote".../>
</sequence>
 </scope>
 <flow>
	<scope>
	<compensationHandler>
		<sequence>
 <invoke partnerLink="BrokerPL"
    operation ="cencelLoan"....../>
  </sequence>
  </compensationHandler>
 <sequence>
<invoke partnerLink="ReqLoan_loanstar",
outputVariable="Reply",
 inputVariable="SupplierQuote"..../>
<reply partnerLink="Reply_broker",
  variable="Reply"...../>
</sequence>
</scope>
<scope>
<compensationHandler>
....
</compensationHandler>
<sequence>
....
</sequence>
</scope>
......
</process>
\end{verbatim}
\end{footnotesize}

\begin{figure}[!htb]
\centering
\includegraphics[scale=.35]{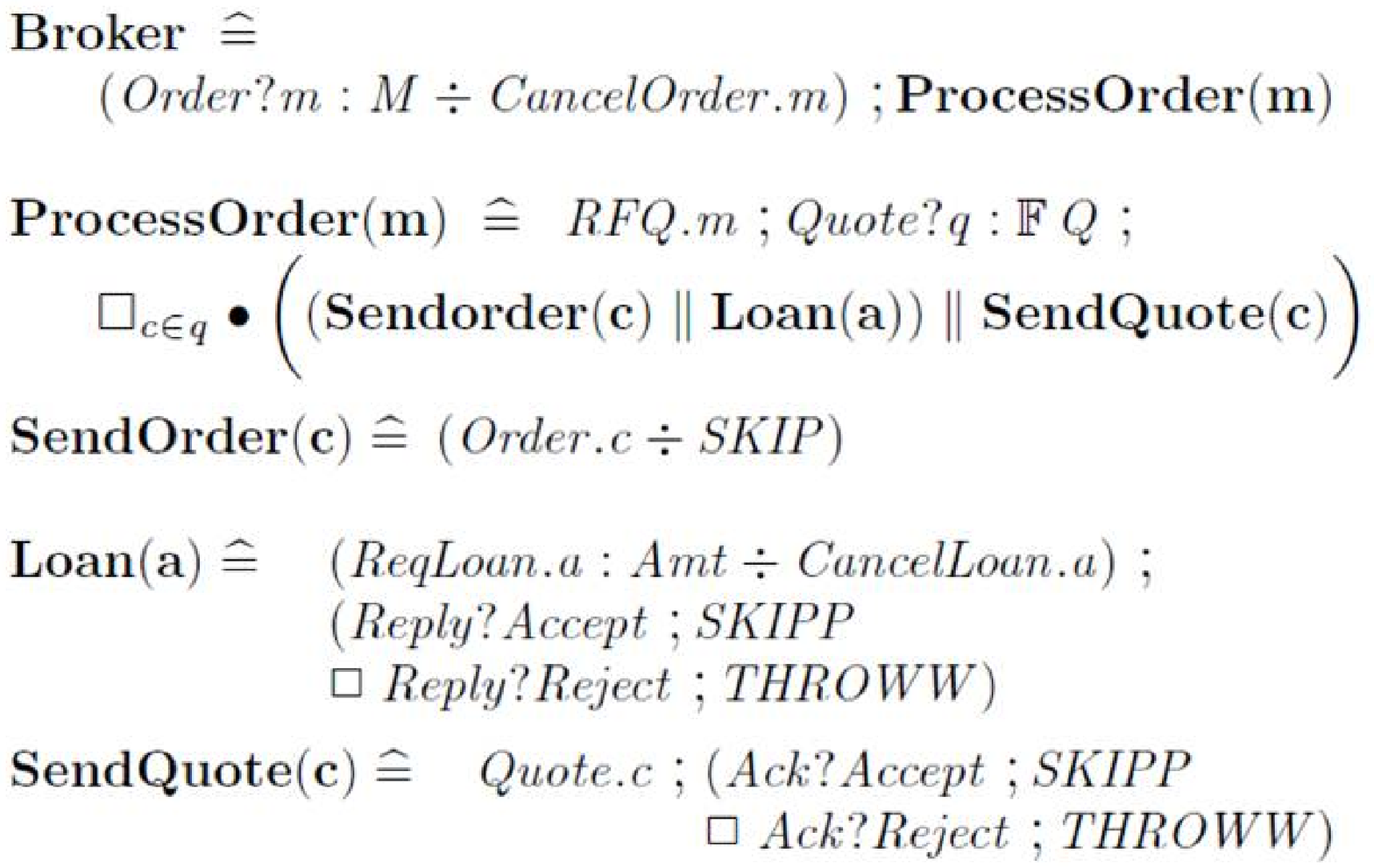}
\caption{cCSP model of Broker service}
\label{fig:brokercsp}
\end{figure}

BPEL construct \verb!sequence! is defined to arrange the services in sequential order, \verb!flow! is used to model the tasks in parallel. Each operation is defined within a scope (\verb!scope!) which is used for the scope of that particular service and the range of compensation handler when an error has occurred and compensation is triggered. Due to brevity, a limited part of the BPEL is presented here. The cCSP representation of the Broker web service is defined in Fig.~\ref{fig:brokercsp}

The first step of the transaction is a compensation pair,
where the primary action is to receive an order from the buyer
and the compensation is to cancel the order. M is used to
represent the finite set of car models ranged over by m.

The Broker requests the Supplier for available quotes
(\textbf{RFQ}) and then selects a quote from the received quotes
(Quote). The Broker arranges a loan for the quoted car by
requesting a loan from \textbf{LoanStar}. The loan amount (Amt) of
loan to be requested is decided from the selected quote and
passed to the process Loan. It requests loan from \textbf{LoanStar}
which is either accepted or rejected. If the loan cannot be
provided then an interrupt is thrown to cancel the actions that
have already taken place. A compensation is added to \textbf{ReqLoan}
(\textbf{CancelLoan}) so that in the case of failure in a later stage the
compensation can be invoked to cancel the event. the quote
is also sent to the buyer (\textbf{SendQuote}). An interrupt can be
raised either by the Buyer by rejecting the quote or by the
\textbf{LoanStar} by rejecting the requested loan. In either case, the
Supplier will terminate yielding an interrupt thrown by the
Broker and compensations from both Broker and Supplier
will run in parallel.

The behaviour of the car broker web service is defined by
combining the behaviour of Broker, Buyer, Supplier, and
LoanStar, where the processes synchronize over the sets A, B
and C.
\begin{eqnarray*}
 {\bf System} &~\defs~& {\bf Buyer}\parallel_{A} \close{\bf Broker\parallel_{B}{\bf Supplier}}\\
 &&\parallel_{C} {\bf LoanStar}
 \end{eqnarray*}
 %   \\[1ex]
$A=\set{~Order,Quote,Ack~}$,\\
$B=\set{~RFQ,Quote,Order,Cancel~}$ \\
$C ~=~ \set{~ReqLoan, Reply~}$

The example illustrates the synchronization of processes
within a transaction block, $\close{\bf Broker\parallel_{B}{\bf Supplier}}$ and between
transaction blocks (\textbf{Buyer} and \textbf{LoanStar} are transaction
blocks). It also outlines how the compensations are handled
in each case.

\subsection{Lender Web Service}

A loan service is a common example of a business process (please refer to~\cite{turner05} for a full description). We assume a lender web service {\bf LoanStar}, that offers loans to online customers. A customer submits a request for an amount to be loaned along with other required information. {\bf LoanStar} first checks the loan amount and if the amount is \pounds10,000 or more, then {\bf LoanStar} asks its business partner {\bf FirstRate} to thoroughly assess the loan. After a detailed assessment of the loan, {\bf FirstRate} can either approve the loan or reject the loan.
A full assessment is costly, so if the loan amount is less than \pounds10,000, the loan is evaluated more simply. {\bf LoanStar} asks its business partner {\bf Assessor} to evaluate the risk for the loan. If the associated risk is low then loan is approved, otherwise {\bf LoanStar} asks {\bf FirstRate} to perform a full assessment.

We give a simple specification of the lender and do not
consider any attached compensation for it. The processes are
defined as standard processes. At the top level, the transaction
is defined as a sequence of two processes: receiving an order
and processing the order.
\begin{eqnarray*}
{\bf LoanStar} &~~\defs~~& LoanOrder?a:Amt \seq {\bf Process}(a)
\end{eqnarray*}

The Process first checks the loan amount to determine the
type of evaluation that needs to be performed. The process
\textbf{ChkAmt} checks the loan amount control is passed to either
Below or Over depending on the loan amount.

\begin{eqnarray*}
 {\bf Process(a)} &~~\defs~~& ChkAmt.a \seq \begin{array}[t]{l}
                          (Below.a\seq {\bf Assessor(a)}\\
                         \extchoice Over.a \seq {\bf FirstRate(a)})
                               \end{array}
\end{eqnarray*}

The process Assessor starts for a loan amount lower than
£10,000. It first checks the risk associated with the loan. If the
risk is low the loan is approved. If the risk is high control is
passed to the \textbf{FirstRate} for a full assessment. After performing
a full assessment and depending on the outcome, \textbf{FirstRate}
either accepts or rejects the requested loan.
\begin{eqnarray*}
 {\bf Assessor(a)} &~~\defs~~& ChkRisk.a \seq \begin{array}[t]{l}
                        (Low.a\seq Reply.Accept \\
                        \extchoice High.a\seq {\bf FirstRate(a)})
                        \end{array}
\\
 {\bf FirstRate(a)} &~~\defs~~& Assess.a \seq \begin{array}[t]{l}
                            (Ok\seq Reply.Accept\\
                        \extchoice NotOk\seq Reply.Reject)
                        \end{array}
\end{eqnarray*}

In the example, we abstract the details of the behaviour of
Assessor and \textbf{FirstRate}. Both of them can be modeled as a
separate web service or as a part of the lender web services. A part of the corresponding BPEL model is as follows:

\begin{footnotesize}
\begin{verbatim}
<process name="LoanStar".../>
<scope>
<compensationHandler>
<sequence>
<invoke partnerLink="LoanStarPL"
  operation ="cencelRequest"....../>
</sequence>
</compensationHandler>
<sequence>
<receive partnerLink="Loan_Req",
  Variable="Amt"...../>
<invoke partnerLink="Chk_AmtPL",
 outputVariable="ProceedLoan",
 inputVariable="Amt"..../>
<reply partnerLink="Amt_Check",
  variable="ProceedLoan"...../>
<invoke partnerLink="BrokerPL",
inputVariable="ConfrimLoan"..../>
...
</sequence>
</scope>
</process>	
\end{verbatim}
\end{footnotesize}

\subsection{Supplier Web Service}

\begin{figure}[!htb]
   \centering
  \includegraphics[scale=.75]{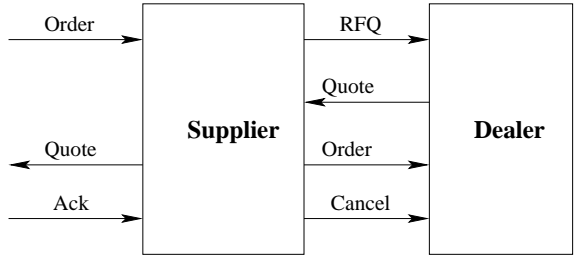}
  \caption{A car supplier web service}
  \label{fig:supplier}
\end{figure}

Car supplier web service provides buyers a good deal on
car orders. Supplier is a supplier that takes orders for a car
from buyers (or from brokers as in Sect.~\ref{sec:broker}). The supplier
sends a request for quotes (RFQ) to a dealer to get available
quotes for the requested car model. The dealer collects quotes
from all of its associated partners and passes the accumulates
quotes to the supplier. An offer (or a list of offers) is selected
by the supplier and sent to the buyer. The offer is also sent
to the dealer as a definite order for the selected model as it is
expected that the buyer will accept the offer. The buyer can
either accept or reject the quote. If the order is rejected by the
buyer, a compensation is invoked to cancel the order that is sent to the dealer. Here, we give a simple representation of the
order receipt and dealer activities and focus on the behaviour
of the car supplier in more detail (Fig.~\ref{fig:supplier}).

\begin{eqnarray*}
  {\bf Supplier} &~~\defs~~& \bigl[~~(Order?m:M\cpair CancelOrder.m)\seq\\
 && {\bf ProcessOrder(m)}~~\bigr]\\
 \mbox{where,}
 && M= \mbox{Car Models}
\end{eqnarray*}

The transaction steps for the \textbf{Supplier} are same as that of
a Broker described earlier, except that lender service is not
attached here.

\begin{eqnarray*}
 {\bf ProcessOrder(m)} &~~\defs~~& RFQ.m \seq RecQuote?q:\power Q\seq\\
  && \Extchoice_{c\in q}\spot
       \biggl((Order.c\cpair Cancel.c)\\
       &&\parallel {\bf SendQuote(c)}\biggr)
    \\
  \mbox{where,}&& Q = \mbox{Available Quotes}
\end{eqnarray*}

After receiving a quote from the supplier, the buyer acknowledges
the receipt of a quote by either accepting or
rejecting it. In the case of rejection, an interrupt is thrown
to compensate the activities that have already taken place. It
has been discussed earlier that as \textbf{SendOrder} and \textbf{SendQuote}
interleave with each other, the interrupt from the buyer can be
thrown before sending the order to the dealer. The compensations
are stored dynamically during the execution of processes,
which is in this case empty and compensation mechanism can
take care of it. The \textbf{SendQuote} process sends a quote, then
receives an acknowledgement as either Accept or Reject.

\begin{eqnarray*}
  {\bf SendQuote(c)} &~~\defs~~& Quote.c \seq \begin{array}[t]{l}
                        (Ack?Accept\seq SKIPP\\
                        \extchoice Ack?Reject\seq THROWW)
                     \end{array}
\end{eqnarray*}

The behaviour of the supplier system is defined by composing
the behaviour of Supplier, Dealer and Buyer.

\begin{eqnarray*}
  {\bf System} &~\defs~& \biggl({\bf Buyer} \parallel_{A} {\bf Supplier}\biggr) \parallel_{B}  {\bf Dealer}\\
  A &=& \set{Order, Quote, Ack}\\
  B &=& \set{RFQ, Quote, Order, Cancel}
\end{eqnarray*}

The BPEL process model is defined as follows,

\begin{footnotesize}
\begin{verbatim}
<process name="Supplier".../>
 <scope>
 <compensationHandler>
  <sequence>
  <invoke partnerLink="SuplierPL"
     operation ="cencelOrder"../>
  </sequence>
 </compensationHandler>
<sequence>
 <receive partnerLink="order_Supplier",
           Variable="orderReq"...../>
 <invoke partnerLink="RFQ_Dealer",
  outputVariable="DealerQuote",
  inputVariable="orderReq"..../>
  <reply partnerLink="Quote_Supplier",
  variable="DealerQuote".../>
 <invoke partnerLink="BrokerPL",
  outputVariable="Ack",
  inputVariable="DealerQuote"..../>
 <reply partnerLink="Reply_Supplier ",
  variable="Ack".../>
 <invoke partnerLink="Order_Dealer",
  outputVariable="confrim",
  inputVariable="DealerrQuote"..../>
  <reply partnerLink="Confrim_Dealer",
  variable="confrim"...../>
 </sequence>
</scope>
</process>		
\end{verbatim}
\end{footnotesize}

\section{Conclusions}

We have shown how cCSP process algebra constructs
can be used to model business transactions. Importantly, we
have shown how compensations are orchestrated to model
the business processes. The compensations are accumulated
during the execution of the processes. The compensations are
defined in such a way that when an interrupt occurs at any
stage of the transaction, the appropriate compensations (which
might be empty when interruption occurs before occurring an
event with attached compensation) are executed for the actions
that already did take place. Having been able to model the web services both by using BPEL and a process algebra confirms suitability of a proper formal verification of the web service composition. As this composition is crucial to business organizations, our comparative case study made a significant leap towards the development of a verified web services. Our future plan includes the encoding the process algebraic model into a suitable model checker such as FDR~\cite{FDR} to check various properties of the whole web service compositions.

%\bibliographystyle{IEEEtran}
%\bibliography{IEEEabrv,ref-all}

\end{document}